\documentclass[aoas,nameyear,seceqn,dvips]{arximspdf}
\usepackage{dcolumn}
\usepackage{stfloats}
\usepackage{graphicx,breakurl}

\doi{10.1214/09-AOAS322}
\volume{4}
\issue{3}
\pubyear{2010}
\firstpage{1342}
\lastpage{1364}

\makeatletter
\newcolumntype{d}[1]{D{.}{.}{#1}}
\renewcommand{\citep}[1]{\citet{#1}}

\def\a{\bolds{\alpha}}
\def\bg{\bolds{\beta_\gamma}}
\def\M{\bolds{\mathcal{M}}}
\def\Mg{\bolds{\mathcal{M}_{\gamma}}}

\fnbelowfloat
\def\S{\bolds{\mathcal{S}}}

\def\tg{\bolds{\theta}_{\bolds{\gamma}}}
\def\tghat{\hat{\bolds{\theta}}_{\bolds{\gamma}}}
\def\Dmiss{G^{\mathrm{m}}}
\def\Dobs{G^{\mathrm{o}}}
\def\SNP{\operatorname{SNP}}
\def\BF{\operatorname{BF}}
\def\z{\mathbf{z}}
\def\xg{\mathbf{x}_{\bolds{\gamma}}}
\def\pg{s_{\bolds{\gamma}}}
\def\R{\texttt{R}}
\newcommand{\Be}{\operatorname{\mathsf{Beta}}}
\newcommand{\BB}{\operatorname{\mathsf{BetaBinomial}}}
\newcommand{\Bin}{\operatorname{\mathsf{Bin}}}

\makeatother

\begin{document}
\begin{frontmatter}

\title{Bayesian model search and multilevel inference for SNP association studies}
\runtitle{Multilevel inference for SNP association studies}

\begin{aug}
\author[A]{\fnms{Melanie A.} \snm{Wilson}\thanksref{t1,t3}\ead[label=e1]{maw27@stat.duke.edu}},
\author[A]{\fnms{Edwin S.} \snm{Iversen}\thanksref{t1,t3}\ead[label=e2]{iversen@stat.duke.edu}},
\author[A]{\fnms{Merlise A.} \snm{Clyde}\corref{}\thanksref{t1,t2}\ead[label=e3]{clyde@stat.duke.edu}},
\author[A]{\fnms{Scott C.} \snm{Schmidler}\ead[label=e4]{scs@stat.duke.edu}} \and
\author[B]{\fnms{Joellen  M.} \snm{Schildkraut}\thanksref{t1,t3}\ead[label=e5]{schil001@mc.duke.edu}}
\runauthor{M. A.~Wilson et al.}
\affiliation{Duke University}
\thankstext{t1}{Supported in part by the NIH Grant NIH/NHLBI R01-HL090559.}
\thankstext{t2}{Supported in part by the  NSF Grants  DMS-03-42172 and DMS-04-06115.  Any opinions, findings and
conclusions or recommendations expressed in this material are those
of the authors and do not  necessarily reflect the views of the
National Science Foundation.}
\thankstext{t3}{Supported by the Duke SPORE in Breast Cancer,
P50-CA068438; the North Carolina Ovarian Cancer Study, R01-CA076016.}

\address[A]{M. A. Clyde\\
E. S. Iversen \\
S. C. Schmidler \\
M. A. Wilson \\
Department of Statistical Science \\
Duke University \\
Durham, North Carolina 27708-0251 \\USA\\
\printead{e3} \\
\phantom{E-mail:} \printead*{e2} \\
\phantom{E-mail:} \printead*{e4} \\
\phantom{E-mail:} \printead*{e1}}

\address[B]{J. M. Schildkraut\\
Department of Community and Family Medicine\\
Duke University\\
Durham, North Carolina 27713\\USA\\
\printead{e5}}
\end{aug}

\received{\smonth{7} \syear{2009}}
\revised{\smonth{12} \syear{2009}}

\begin{abstract}
Technological advances in genotyping have given rise to
hypothesis-based association studies of increasing scope. As a
result, the scientific hypotheses addressed by these studies have
become more complex and more difficult to address using existing
analytic methodologies.  Obstacles to analysis include inference in
the face of multiple comparisons, complications arising from
correlations among the SNPs (single nucleotide polymorphisms),
choice of their genetic parametrization and missing data.  In this
paper we present an efficient Bayesian model search strategy that
searches over the space of genetic markers and their genetic
parametrization.  The resulting method for Multilevel Inference of
SNP Associations, MISA, allows computation of multilevel posterior
probabilities and Bayes factors at the global, gene and SNP level,
with the prior distribution on SNP inclusion in the model providing
an intrinsic multiplicity correction.   We use
simulated data sets to characterize MISA's statistical
power, and show that MISA  has higher power to detect association than
standard procedures. Using data from the North
Carolina Ovarian Cancer Study (NCOCS), MISA identifies variants that were
not identified by standard methods and have been externally
``validated''
in independent studies.  We examine sensitivity of the
NCOCS results to prior choice and method for imputing missing data.
MISA is available in an \R\ package on CRAN.
\end{abstract}

\begin{keyword}
\kwd{AIC}
\kwd{Bayes factor}
\kwd{Bayesian model averaging}
\kwd{BIC}
\kwd{Evolutionary Monte Carlo}
\kwd{false discovery}
\kwd{genetic models}
\kwd{lasso}
\kwd{model uncertainty}
\kwd{single nucleotide polymorphism}
\kwd{variable selection}.
\end{keyword}

\end{frontmatter}

\setcounter{footnote}{3}

\section{Introduction}
Recent advances in genotyping technology have resulted in a dramatic
change in the way hypothesis-based genetic association studies are
conducted.  While previously investigators were limited by costs to
investigating only a handful of variants within the most interesting
genes, researchers may now conduct candidate-gene and
candidate-pathway studies that encompass many hundreds or thousands
of genetic variants, often single nucleotide polymorphisms (SNPs).
For example, the North Carolina Ovarian Cancer Study
(NCOCS) [\citet{Schildkraut2008}], an ongoing population-based case-control
study, genotyped $2129$ women at $1536$ SNPS in $170$
genes on $8$ pathways, where ``pathway'' is defined as a set of genes
thought to be simultaneously active in certain circumstances.

The analytic procedure most commonly applied to association studies of
this scale is to fit a separate model of association for each SNP that
adjusts for design and confounder variables.  As false discoveries due
to multiple testing are often a concern, the level of significance for
each marginal test of association is adjusted using Bonferroni or
other forms of false discovery correction
[\citep{Storey2002}; \citep{ProbFalseFind2004}; \citep{Bald2006}].  While these methods
have been shown to be effective in controlling the number of false
discoveries reported, correlations between the markers may limit the
power to detect true associations [\citep{Efron2007}].  The NCOCS study
provides a case in point.  When simple marginal methods are applied
to the NCOCS data, no SNPs are identified as notable.

Marginal SNP-at-a-time methods do not address directly many of the
scientific questions in candidate pathway studies, such as  ``Is there an
overall association between a pathway and the outcome of interest?''
and ``Which genes are most likely to be
driving this association?''  The Multilevel Inference for SNP
Association (MISA) method we describe here is designed to
simultaneously address these questions of association at the level of SNP,
gene and pathway.

MISA, in contrast to the marginal methods, identifies ten SNPs of
interest in the NCOCS study.  To date, one of these (ranked tenth by
MISA) has been validated in external data by a large multi-center
consortium [\citep{Schietal2009}]; additional testing is underway for
other top SNPs discovered by MISA.  To buttress this empirical
evidence, we demonstrate using simulation studies (Section~\ref{section:compare}) that MISA has higher power to detect
associations than other simpler procedures, with a modest increase in
the false discovery rate (Figure~\ref{power.ps}).

In the next section we describe the Bayesian hierarchical model
behind MISA and highlight how it addresses many of the key issues in
analysis of SNP association studies: identification of associated SNPs
and genetic models, missing data, inference for multi-level
hypotheses and control of the false discovery rate.  Like stepwise
logistic regression [\citep{Bald2006}], lasso
[\citep{ParkHast2008}; \citep{Shietal2007}; \citep{Wuetal2009}] and logic regression
[\citep{2003logic}; \citep{mcmclogic}; \citep{KatjaLogicReg}], MISA improves upon
marginal, SNP-at-a-time methods by modeling the outcome variable as
a function of a multivariate genetic profile, which provides measures
of association that are adjusted for the remaining markers.  MISA uses
Bayesian Model Averaging [\citep{HoetMadiRaftVoli1999}] to combine
information from multiple models of association to address the degree
to which the data support an association at the level of individual
SNPs, genes and pathways, while taking into account uncertainty
regarding the best genetic parametrization.  By using model averaging,
MISA improves upon methods that select a single model, which may miss
important SNPs because of LD structure. We show how the prior
distribution on SNP inclusion provides a built-in multiplicity
correction.  Because missing data are a common phenomenon in
association studies, we discuss two options for handling this problem.

In Section~\ref{section:modelsearch} we present an Evolutionary Monte
Carlo algorithm to efficiently sample models of association according
to their posterior probabilities.  In Section~\ref{section:compare} we apply our method to
simulated data sets and demonstrate that MISA outperforms less complex
and more commonly used alternatives for detecting associations in
modestly powered candidate-gene case-control studies.  The
simulation approach may also be used to guide selection of the prior
hyperparameters given the study design.  In Section~\ref{sec:NCOCS} we return to
the NCOCS study and present results from the analysis of a single
pathway from that study.  We examine the sensitivity of results to
prior hyperparameter choice and methods for imputing missing data.  We
conclude in Section~\ref{sec6} with recommendations and a discussion of future
extensions.

\section{Models of association}\label{section:modelspec}
We consider SNP association models with a binary phenotype, such as
presence or absence of a disease as in case-control designs.  For
$i=1, \ldots, n$, let $D_i$ indicate the disease status of individual
$i$, where $D_i = 1$ represents a disease case and $D_i = 0$
represents a control.  For each individual, we have $S$ SNP
measurements, where SNP $s$ is either homozygous common ($A_sA_s$),
heterozygous ($a_sA_s$ or $A_s a_s$), homozygous rare ($a_s a_s$), or
missing and is coded as 0, 1, 2, representing the number of rare
alleles, or NA if the SNP is missing for that individual.  We will
discuss methods for imputing missing SNP data in Section~\ref{sec:missingdata}.  In addition to the SNP data, for each
individual we have a $q$-dimensional vector $\z_i^{T}$ of design and
potential confounding variables that will be included in all models,
henceforth referred to as ``design'' variables.

We use logistic regression models to relate disease status to the
design variables and subsets of SNPs. We denote the collection of all
possible models by $\M$. An individual model, denoted by $\Mg$, is
specified by the $S$ dimensional vector $\gamma$, where
$\gamma_s$ indicates the inclusion and SNP-specific genetic
parametrization of SNP $s$ in model $\Mg\dvtx  \gamma_s = 0$ if $\SNP_s
\notin \Mg$, $\gamma_s = 1$ if $\SNP_s \in \Mg$ with a log-additive
parametrization, $\gamma_s = 2$ if $\SNP_s \in \Mg$ with a dominant
parametrization, and $\gamma_s = 3$ if $\SNP_s \in \Mg$ with a
recessive parametrization.  When no homozygous rare cases or controls
are observed, we fix the genetic parametrization to be log-additive.
Under each of these genetic parametrizations, SNP $s$ may be encoded
using one degree of freedom. In particular, for the log-additive
model, the design variable representing SNP $s$ is a numeric variable
equal to the number of copies of the risk allele $a_s$.  For the
dominant model, we use an indicator variable of whether allele $a_s$
is present (homozygous rare or heterozygous) and for the recessive
model, an indicator variable of whether SNP $s$ has the homozygous
rare genotype. For each individual, the logistic regression under
model $\Mg$ assuming complete data is given by
\begin{equation}\label{eq:model}
\operatorname{logit}\bigl(p(D_i = 1|\z_i,{\xg}_i,\tg,\Mg)\bigr)
=  \alpha_0 + \z_i^{T} \a + {\mathbf{x}^{T}_{\bolds{\gamma}}}_i \bg,
\end{equation}
where ${\xg}_i$ represents the coding of SNPs in model $\Mg$ and $\tg$
is the vector of model specific parameters $(\alpha_0, \a^T, \bolds{\beta}^T_{\bolds{\gamma}})$,
with intercept $\alpha_0$, vector of design variable coefficients
$\a$ and log-odds ratios $\bg$.  Prospective models for disease
outcome given multivariate genetic marker data as in equation
(\ref{eq:model}) provide measures of association that are adjusted for
other markers which can increase the power to detect associations
[\citep{Bald2006}], however, one is faced with an extremely large
collection of possible models. While stepwise selection methods may be
used to select a single model [\citep{CordClay2002}], this leads to
difficulty in interpreting the significance of SNPs in the selected
model.  Bayesian model averaging is an alternative to stepwise
selection methods and is an effective approach for identifying subsets
of likely associated variables, for prioritizing them and for
measuring overall association in the presence of model uncertainty
[see the review articles by \citet{HoetMadiRaftVoli1999} and
\citet{ClydGeor2004} and the references therein].

\subsection{Posterior inference}
Posterior model probabilities measure the degree to which the data
support each model in a set of competing models.  The posterior model
probability of any model $\Mg$ in the space of models $\M$ is
expressed as
\begin{eqnarray*}
p(\Mg \vert D) = \frac{p(D \vert \Mg) p(\Mg)}{\sum_{\Mg \in \M}
p(D \vert \Mg) p(\Mg )} \qquad\mbox{for }  \Mg \in \M,
\end{eqnarray*}
where $p(D \vert \Mg)$ is the (marginal) likelihood of model $\Mg$
obtained after integrating out model-specific parameters $\tg$ with
respect to their prior distribution, and $p(\Mg )$ is the prior
probability of $\Mg$.

While posterior probabilities provide a measure of evidence for
hypotheses or models, it is often difficult to judge them in isolation,
as individual model probabilities may be ``diluted'' as the space of
models grows [\citep{Clyd1999}; \citep{Geor1999}; \citep{ClydGeor2004}].  Bayes
factors (BF) [\citep{KassRaft1995}] compare the posterior odds of any
two models (or hypotheses) to their prior odds
\[
\BF(\Mg_1\dvtx \Mg_2)
= \frac{p(\Mg_1 \vert D)/p(\Mg_2 \vert  D)}{p(\Mg_1)/p(\Mg_2)}
\]
and measures the \textit{change} in evidence (on the log scale) provided by
data for one model, $\Mg_1$, to another, $\Mg_2$, or for pairs of
hypotheses.  \citet{Goodman1999}  and
\citet{StepBald2009} provide a discussion on the
usefulness of Bayes factors in the medical context and
\citet{WakefieldBFDP2007} illustrates their use in controlling false
discoveries in genetic epidemiology studies.  Below we define Bayes
factors for quantifying association at multiple levels (global, gene
and SNP) and assessing the most likely SNP-specific genetic
parametrization.

\subsubsection{Global Bayes factor}
The Bayes factor in favor of $H_A$, the alternative hypothesis that
there is at least one SNP associated with disease, to $H_0$, the null
hypothesis that there is no association between the SNPs under
consideration and disease, measures the relative weight of evidence of
$H_A$ to $H_0$.  The null model corresponding to $H_0$ is the model
which includes only design variables and no SNPs, and is denoted
$\M_0$.  The alternative hypothesis is represented by all of the
remaining models in $\M$.  Because the space of models is large, the
null model (or any single model in general) may receive small
probability (both prior and posterior), even when it is the highest
posterior probability model (this illustrates the dilution effect of
large model spaces); Bayes factors allow one to judge how the
posterior odds compare to one's prior odds.

The Global Bayes factor for comparing $H_A$ to $H_0$ may be simplified to
\begin{eqnarray}\label{eq:globalBF}
\BF(H_A \dvtx  H_0)  = \sum_{\Mg \in \M} \BF(\Mg \dvtx  \M_0) p(\Mg  \vert  H_A),
\end{eqnarray}
which is the weighted average of the individual Bayes factors
$\BF(\Mg \dvtx  \M_0)$ for comparing each model in $H_A$ to the null model with
weights given by the prior probability of $\Mg$ conditional on being
in $H_A$, $p(\Mg \vert H_A)$. Because the alternative is a composite
hypothesis, the resulting Global Bayes factor is not independent of the prior
distribution on the models that comprise the alternative, thus, the
prior distribution on models will play an important role in
controlling the (relative) weights that models of different sizes receive.
For a large number of SNPs, it is
impossible to enumerate the space of models and posterior summaries
are often based on models sampled from the posterior distribution. In
equation (\ref{eq:globalBF}), if we replace the average over all
models in $H_A$ with the average over the models in $\S$ (the
collection of unique models sampled from the posterior distribution),
the result
\begin{eqnarray*}
\BF(H_A\dvtx  H_0)  >  \BF_S(H_A \dvtx  H_0) \equiv \sum_{\Mg \in \S}   \BF(\Mg\dvtx  \M_0) p(\Mg  \vert  H_A)
\end{eqnarray*}
is a lower bound for the Bayes factor for testing global association.
If the lower bound indicates evidence of an association, then we can
be confident that this evidence will only increase as we include more
models.

\subsubsection{SNP  Bayes factors}
While it is of interest to quantify association at the global level,
interest is primarily in identifying the gene(s) and variant(s) within
those genes that drive the association.  We begin by defining SNP
inclusion probabilities and associated Bayes factors.  These marginal
summaries are adjusted for the other potentially important SNPs and
confounding variables and provide a measure of the strength of
association at the level of individual SNPs.  Given each sampled model
$\Mg \in \S$ and the model specification vectors $\gamma = (\gamma_1,
\gamma_2,\ldots, \gamma_S)$ previously defined in Section~\ref{section:modelspec}, the inclusion probability for SNP $s$ is
estimated as
\begin{equation} \label{eq:SNP-prob}
p(\gamma_s \neq 0 \vert D)  = \sum_{\Mg \in \S} 1_{ (\gamma_s \ne 0)} p(\Mg  \vert  D, \S),
\end{equation}
where $p(\Mg \vert D, \S)$ is the posterior probability of a model
re-normalized over the sampled model space.  The SNP Bayes factor is
the ratio of the posterior odds of the SNP being associated to the
prior odds of the same, and is defined as
\begin{eqnarray*}
\BF(\gamma_s\neq0\dvtx \gamma_s=0)  = \frac{p(\gamma_s\neq0 \vert
D)}{p(\gamma_s=0 \vert
D)}\div \frac{p(\gamma_s\neq0)}{p(\gamma_s=0)},
\end{eqnarray*}
where $p(\gamma_s \neq 0)$ is the prior probability of SNP $s$ being
associated.  Estimates of the SNP Bayes factor may be obtained using
the estimated SNP inclusion probabilities from (\ref{eq:SNP-prob}).

\subsubsection{Gene Bayes factors}
In cases where there are SNPs in Linkage Disequilibrium (LD), SNP
inclusion probabilities may underestimate the significance of an
association at a given locus.  This occurs because SNPs in LD may
provide competing explanations for the association, thereby diluting
or distributing the probability over several markers.  Since the
amount of correlation between markers across different genes is
typically negligible, calculating inclusion probabilities and Bayes
factors at the gene level will not be as sensitive to this dilution.
A gene is defined to be associated if one or more of the SNPs within
the given gene are associated. Hence, we define the gene inclusion
probability as
\begin{eqnarray*}
p(\Gamma_g = 1 \vert D)  = \sum_{\Mg \in \S} 1_{ (\Gamma_{g} = 1)} p(\Mg  \vert  D, \S),
\end{eqnarray*}
where $\Gamma_{g} = 1$ if at least one SNP in gene $g$ is in model
$\Mg$ and is zero otherwise.  The gene Bayes factor is defined as
\begin{eqnarray*}
\BF(\Gamma_{g}=1\dvtx \Gamma_{g}=0)  = \frac{p(\Gamma_{g}=1 \vert
D)}{p(\Gamma_{g}=0 \vert
D)}\div \frac{p(\Gamma_{g}=1)}{p(\Gamma_{g}=0)},
\end{eqnarray*}
where $p(\Gamma_{g} = 1)$ is the prior probability of one or more SNPs
in gene $g$ being associated.

\subsubsection{Interpreting evidence}\label{sec2.1.4}

Jeffreys [(\citeyear{Jeff1961}), page 432], presents a descriptive classification of
Bayes factors into ``grades of evidence'' (reproduced in Table~\ref{table:jeff}) to assist in their interpretation
[see \citet{KassRaft1995}].  In the context in which he presents the
grades, he defined the Bayes factor assuming equal prior odds, making
it equivalent to posterior odds and enabling a meaningful
interpretation in terms of probabilities.  It is not clear whether he
intended his descriptive grades to be used more broadly for
interpreting Bayes factors or for interpreting posterior
probabilities.

Jeffreys was well aware of the issues that arise with testing several
simple alternative hypotheses against a null hypothesis
[\citeauthor{Jeff1961} (\citeyear{Jeff1961}), Section~5.04], noting that if one were to test several hypotheses
separately, that by chance one might find one of the Bayes factors to
be less than one even if all null hypotheses were true.  He suggested
that, in this context, the Bayes factors needed to be ``corrected for
selection of hypotheses'' by multiplying by the prior odds.

Experience has shown that detectable SNP associations are relatively
infrequent, hence, the prior odds of any given SNP being marginally
associated in the typical genetic association study should be small.
For this reason, \citet{StepBald2009} suggest that marginal Bayes
factors calculated assuming equal prior odds be interpreted in light
of a prior odds more appropriate to the study at hand.  Our approach
to the problem of exploring multiple hypotheses is to embed each of
the potential submodels (corresponding to a subset of SNPs) into a
single hierarchical model.  Unlike the marginal (one-at-a-time) Bayes
factors in \citet{StepBald2009} that are independent of the prior
odds on the hypotheses, our SNP Bayes factors are based on comparing
composite hypotheses and hence do depend on the prior distribution
over models, which implicitly adjusts for the selection of hypotheses.

While Bayes factors do not provide a measure of \textit{absolute} support
for or against a hypothesis (except with even prior odds), the log
Bayes factor does provide a coherent measure of how much the data \textit{change} the support for the hypothesis (relative to the prior)
[\citep{LaviSche1997}]. Applying Jeffreys grades to Bayes factors
using priors distributions that account for competing hypotheses
provides an idea of the impact of the data on changing prior beliefs,
but ultimately posterior odds provide a more informative measure of
evidence and model uncertainty.

\begin{table}
\caption{Jeffrey's grades of evidence [\protect\citep{Jeff1961}, page 432]}\label{table:jeff}
\begin{tabular*}{9cm}{@{\extracolsep{4in minus 4in}}llc@{}}
\hline
\textbf{Grade} & \multicolumn{1}{c}{$\operatorname{\mathbf{BF}}\bolds{(H_A\dvtx H_0)}$} & \textbf{Evidence against} $\bolds{H_0}$\\
\hline
1 & \phantom{3.6}$1$ -- $3.2$ & Indeterminate \\
2 & \phantom{3}$3.2$ -- $10$ & Positive  \\
3 & \phantom{.6}$10$ -- $31.6$ & Strong  \\
4 & $31.6$ -- $100$ & Very strong  \\
5 & \multicolumn{1}{c}{$>$100} & Decisive \\
\hline
\end{tabular*}
\end{table}

\subsection{Prior distributions, Laplace approximations and marginal likelihoods}
We assume normal prior distributions for the coefficients $\tg$ with a
covariance matrix that is given by a constant $1/k$ times the inverse
Fisher Information matrix. For logistic regression models, analytic
expressions for $p(D \vert \Mg)$ are not available and Laplace
approximations or the Bayes Information Criterion are commonly used to
approximate the marginal likelihood
[\citep{Raft1986}; \citep{WakefieldBFDP2007}; \citep{WellcomeTrust2007}].  Using a
Laplace approximation with the normal prior distribution [\citep{Wilsonetal2010}], the posterior probability of model $\Mg$ takes the form of a
penalized likelihood
\begin{equation} \label{eq:post-prob}
p(\Mg \vert D) \propto \exp\bigl\{- \tfrac{1}{2}[\operatorname{dev}(\Mg; D)+
\operatorname{pen}(\Mg)]\bigr\},
\end{equation}
where $\operatorname{dev}(\Mg; D)= - 2 \log(p(D \vert \tghat,\Mg))$ is the
model deviance, and the
penalty term $\operatorname{pen}(\Mg)$ encompasses a penalty on model size
induced by the choice of $k$ in the prior distribution on
coefficients $\tg$ and the prior distribution over models.  Because we
expect that  effect sizes will be small, we calibrate the
choice of $k$ based on the Akaike information criterion
[\citep{Wilsonetal2010}], leading to
\[
\operatorname{pen}(\Mg) = 2(1 + q + \pg) - 2 \log(p(\Mg)).
\]

\subsection{Missing data}\label{sec:missingdata} The expression in
(\ref{eq:post-prob}) assumes complete data on all SNPs. Missing SNP
data, unfortunately, are the norm rather than the exception in
association studies. Removing all subjects with any missing SNP
genotype data will typically result in an unnecessary loss of
information and potential bias of estimated effects if the missing
data are nonignorable. It is possible, however, to exploit patterns
in LD to efficiently impute the missing genotypes given observed data
[\citep{Bald2006}].  We use fastPHASE [\citep{Phase2001}; \citep{ServStep2007}]
to sample haplotypes and missing genotypes ($\Dmiss$) given the
observed unphased genotypes ($\Dobs$).  This assumes that the pattern
of missing data is independent of case-control status, which, if not
true, may lead to serious biases [\citep{Clayetal2005}]. This
assumption may be examined  by using indicator variables of
missingness as predictors in MISA.

The posterior probabilities of models given the data are obtained by
averaging the marginal likelihood of a model over imputed genotype
data:
\begin{eqnarray} \label{eq:post-prob-miss}
p(\Mg \vert D)
&\propto& \int \exp\bigl\{- \tfrac{1}{2}[\operatorname{dev}(\Mg; D, \Dobs,\Dmiss)\nonumber
\\
&&\hspace*{91pt}{}+ \operatorname{pen}(\Mg)]\bigr\} p(\Dmiss \vert \Dobs)\, d\Dmiss  \nonumber \\[-8pt]\\[-8pt]
&\approx& \frac{1}{M} \sum^I_{i = 1} \exp\bigl\{- \tfrac{1}{2}[\operatorname{dev}(\Mg; \Dobs, \Dmiss_i)+ \operatorname{pen}(\Mg)]\bigr\}\nonumber
\\
&\equiv& \Psi(\Mg),\nonumber
\end{eqnarray}
where $I$ is the number of imputed data sets, $\operatorname{dev}(\Mg; D,
\Dobs, \Dmiss)$ is the deviance based on the completed data, and
$\Psi(\Mg)$ is an estimate of the un-normalized posterior model
probability for model $\Mg$.  We have found that the number of imputed
sets must be on the order of $I = 100$ to provide accurate estimates
of posterior quantities.  This has a significant computational impact
in the model search algorithm described in Section~\ref{section:modelsearch}.  As a simple alternative, we approximate
(\ref{eq:post-prob-miss}) by a modal approximation, where the missing
genotypes are imputed with the mode of the sampled genotypes using
fastPHASE. While it is well known that plugging in a single estimate
for the missing data under-estimates uncertainty, the modal
approximation provides dramatic computational savings.  In Section~\ref{sec:NCOCS} we examine the sensitivity of results to the method of
imputing missing data and find that the modal approximation gives
comparable results for SNP BFs.

\subsection{Choice of prior distribution on models} \label{ssec:prior}
The prior distribution on the space of models $\M$, $p(\Mg)$,
completes our model specification.  The frequentist approach for SNP
association studies usually involves some form of adjustment for
multiple-testing, which can, in effect, penalize the researcher who
looks beyond single-SNP models of association to multiple SNP models
or models of interactions.  Under the Bayesian approach, posterior
evidence in the data is judged against the prior odds of an
association using Bayes factors, which should not be affected by the
number of tests that an investigator chooses to carry out
[\citep{Bald2006}].

While it has been common practice to adopt a ``noninformative''
uniform distribution over the space of models for association (this is
after marginalizing over the possible genetic models for each SNP),
this choice has the potentially undesirable ``informative''
implication that $\frac{1}{2}$ of the SNPs are expected to be
associated a priori, and the prior odds of at least one SNP
being included (which is used in the global Bayes factor) depends on
the number of tests ($2^S$) (Table~\ref{prior.dilute}).

\begin{table}[b]
\caption{General prior characteristics and limiting behavior (in
parentheses) of the
$\Bin(S, 1/2)$, $\BB(1,1)$ and $\BB(1,\lambda S)$ distribution on model size}\label{prior.dilute}
\begin{tabular*}{\textwidth}{@{\extracolsep{4in minus 4in}}l ccc@{}}
\hline
& \textbf{Binomial}   & \textbf{Beta-Binomial} & \textbf{Beta-Binomial}\\
&  $\bolds{(S, 1/2)}$            & $\bolds{(1, 1)}$         & $\bolds{(1, \lambda S)}$ \\
\hline
Expected model size& $\frac{S}{2}$ ($\infty$) & $\frac{S}{2}$ ($\infty$) & $\frac{S}{\lambda S +1}$ ($\frac{1}{\lambda}$)\\[3pt]
Global prior odds & $\frac{2^{2S}}{2^S+1}$ ($\infty$) & $S$ ($\infty$) & $\frac{1}{\lambda}$ \\
\quad of an association  & & & \\
Marginal prior odds & 1 & 1 & $\frac{1}{\lambda S}$ ($0$) \\
\quad of an association  & & & \\
Prior odds of adding a variable   & $1$ & $\frac{\pg+1}{S -\pg}$ ($0$) & $ \frac{\pg + 1}{(\lambda+1)S - \pg - 1}$ ($0$) \\
\hline
\end{tabular*}
\end{table}

A recommended alternative is the Beta-Binomial distribution on the
model size, which provides over-dispersion, added robustness to prior
misspecification and multiplicity corrections as a function of the
number of variables [\citep{BB2007}; \citep{ScottBerger2008}; \citep{CuiGeorge2008}].
We construct a hierarchical prior distribution over the space of
models defined by subsets of SNPs and their genetic parametrizations
as follows.  For any SNP included in the model, we assign a uniform
distribution over the possible genetic parametrizations.  The prior
distribution on the model size $\pg$ is $\Bin(S, \rho)$ conditional on
$\rho$, and for the last stage, $\rho$ is assigned a $\Be(a,b)$
distribution.  Integrating over the distribution on $\rho$ leads to
the $\BB(a,b)$ distribution on model size,
\begin{equation}
\label{eq:BB}
p(\pg) =  \frac{B(\pg + a, S - \pg + b)}
{(S+1) B(\pg+1,S - \pg + 1)B(a,b)},
\end{equation}
and the following distribution on models,
\begin{equation}
\label{eq:BB-model-prob}
p(\Mg) =  \biggl(\frac{1}{3} \biggr)^{\pg} \frac{B(\pg + a, S - \pg +
b) }{B(a,b) },
\end{equation}
where $B(\cdot, \cdot)$ is the beta function and the factor of $1/3$
accounts for the distribution over genetic parametrizations.

\subsubsection{Default hyperparameter choice}\label{section:hyperchoice}
Following \citet{BB2007} and \citet{ScottBerger2008}, we recommend $a
= 1$ as a default, so that the prior distribution on model size is
nonincreasing in $\pg$.  The hyperparameter $b$ can then be chosen to
reflect the expected model size, the global prior probability of at least
one association or the marginal prior odds that any SNP is associated
(Table~\ref{prior.dilute}).
A default choice is to set $b=1$, leading to a uniform distribution on
model size [\citep{BB2007}; \citep{ScottBerger2008}]. Like the binomial
distribution, the $\BB(1,1)$ distribution results in an expected model
size of $\frac{S}{2}$ (Table~\ref{prior.dilute}), although the
$\BB(1,1)$ distribution has a larger variance than the $\Bin(S, 1/2)$.
Alternatively, if $b$ is proportional to $S$, $b= \lambda S$, the
expected model size approaches a limit of $\frac{1}{\lambda}$ as $S$
approaches infinity.

The choices for hyperparameters have implications for the global Bayes
factor.  The $\BB(1,1)$ has a global prior odds of association equal
to the number of SNPs, $S$, and would be appropriate for the case
where increasing the number of SNPs under consideration reflects
increased prior certainty that an overall (global) association can be
detected.  Under the $\BB(1,\lambda S)$, the global prior odds are
constant, $1/\lambda$, reflecting a prior odds for overall association
that is independent of the number of genes/SNPs tagged.  Also, with
both Beta--Binomial prior distributions, the prior odds of
incorporating an additional SNP in any model decreases with model size
$\pg$ and approaches $0$ in the limiting case as the number of SNPs, $S$,
increases.  This provides an implicit multiple testing correction in the
number of SNPs (rather than tests) that are included in the study of
interest.  The $\BB(1,\lambda S)$ achieves this by keeping the global
(pathway) prior odds of an association constant while decreasing the
marginal prior odds of any one of the SNPs being associated as the
number of SNPs increases.  As a skeptical ``default'' prior, we
suggest the hyperparameters $a=1$ and $b= S$ which leads to the
global prior odds of there being at least one association of $1$ and
the marginal prior odds of any single SNP being associated of $1/S$.


\section{Stochastic search for SNPs}\label{section:modelsearch}
Given the number of SNPs under consideration, enumeration of all
models for $S$ greater than 25--30 is intractable.  While it is
possible to enumerate all single variable SNP models, the number of
models with 2 or 3 SNPs allowing for multiple genetic parametrizations
is in the millions or more for a typical modern hypothesis-oriented
study.  Stochastic variable selection algorithms [see \citet{ClydGeor2004}, for a review] provide a more robust search procedure than
stepwise methods, but also permit calculation of posterior
probabilities and Bayes factors based on a sample of the most likely
candidate models from the posterior distribution.

MISA makes use of a stochastic search algorithm based on the
Evolutionary Monte Carlo (EMC) algorithm of \citet{EMC2000}. EMC is a
combination of parallel tempering [\citep{ParallelTemp1991}] and a
genetic algorithm [\citep{GeneticAlgor1975}] and samples models based on
their ``fitness.'' While originally designed to find optimal models
based on AIC, in our application the fitness of the models is given by
$\psi(\Mg)$,
\begin{eqnarray*}
\psi(\Mg) = \log(\Psi(\Mg)),
\end{eqnarray*}
where $\Psi(\Mg)$ is defined in  equation
(\ref{eq:post-prob-miss}) and is equal  to the log of the un-normalized
posterior model probability.  This results in models being generated
according to their posterior probability.

The EMC algorithm requires that we specify the number of parallel
chains that are run and the associated temperature for each chain that
determines the degree of annealing.  If the temperatures are too
spread out for the number of chains, then the algorithm may exhibit
poor mixing and slow convergence.  \citet{EMC2000} show that even with
all chains run at a temperature of 1 (no annealing), EMC outperforms
alternative sampling methods such as Gibbs sampling and Reversible
Jump MCMC in problems where strong correlations among the predictor
variables lead to problems with exploring multiple modes in the
posterior distribution.  We have found that a constant temperature
ladder with 5 parallel chains provides good mixing and finds more
unique models than using a custom temperature ladder based on the
prescription in \citet{EMC2000}, and recommend the constant
temperature ladder as a default. To assess convergence, we take two
independent EMC runs using randomly chosen starting points and examine
trace plots of the fitness function. We use the marginal likelihoods
from the set of unique models in the sample for inference and compute
estimates of marginal posterior inclusion probabilities for each run.
We continue running the two instances of the EMC algorithm until the
posterior probabilities derived from each are sufficiently close.
This leads to longer running times than those suggested by
conventional convergence diagnostic such as Gelman--Rubin
[\citep{GelmanRubin1992}].

Efficiency of stochastic algorithms often diminishes as the total
number of models increases. For this reason, we have found it useful
to reduce the number of SNPs included in the EMC search using a screen
when $S$ is large.  Such a screen will typically be fairly permissive,
leaving only the weakest candidates out of the stochastic search.  The
screen should be quick to calculate, adjust for the same design
variables and consider the same genetic parametrizations as in the
full analysis.  In our analyses, we calculated marginal (i.e.,
SNP-at-a-time) Bayes factors for each of the log-additive, dominant
and recessive models of association against the model of no
association.  We ordered SNPs according to the maximum of the three
marginal Bayes factors and retained those with a maximum marginal BF
greater than or equal to one.  More details are available in
\citet{Wilsonetal2010}.

\section{Simulation comparison}\label{section:compare}
We used the 124 simulated case-control data sets [details of the
simulation can be found in \citet{Wilsonetal2010}] to estimate true
and false positive
rates for MISA and seven other alternative procedures:
\begin{longlist}
\item[\textit{Bonferroni}] We fit a  logistic regression
model for each SNP under the log-additive parametrization and
calculate the $p$-value for testing association using
a Chi-Squared test.  We use a
Bonferroni corrected level $\alpha=0.05$ test to declare a SNP
associated.
\item[\textit{Adjusted Bonferroni}] We fit a  logistic regression
model for each SNP under the log-additive parametrization and
calculate the $p$-value for testing association using
a Chi-Squared test.  We use a
Bonferroni corrected level $\alpha$ test to declare a SNP
associated where $\alpha$ is chosen so that the proportion of false positives
detected is the same as in MISA
using the default $\BB(1,S)$ prior.
\item[\textit{Benjamini--Hochberg}] We fit the same SNP-at a time logistic
regression as above, but declare a SNP to be associated if it has a
Benjamini--Hochberg false discovery rate of less than $0.05$.
\item[\textit{Marginal BF}] This also utilizes the single SNP at a time
logistic regression, but calculates a BF for association under each
of the three genetic models.  If the maximum BF over the three
genetic models is greater than $3.2$, we declare the SNP associated.
See \citet{Wilsonetal2010} for more detail.
\item[\textit{Stepwise LR (AIC)}] We use a stepwise multiple logistic
regression procedure to select SNPs based on AIC. Each SNP is coded
using 2 degrees of freedom to select among the three genetic models.
SNPs in the final model are called associated.
\item[\textit{Stepwise LR (BIC)}] Same as above but using BIC to select models.
\item[\textit{Lasso}] We use the \texttt{Lasso2} package in \texttt{R} [\citep{lasso2}]
that is based on the algorithm developed by \citet{Osbor1999} to
select SNPs based on the least absolute shrinkage and selection
operator. Each SNP is coded using 2 degrees of freedom to represent
the three genetic models and all SNPs in the final model with
coefficients greater than zero are called associated.
\item[\textit{MISA}] We reduced the number of SNPs using the marginal
Bayes factor method above to eliminate SNPs with a marginal BF $\ge
1$. We ran MISA using the default
$\BB(1,S)$ and the $\BB(1/8,S)$ prior distributions
on the models using two runs of 400,000 iterations based on convergence
of the marginal inclusion probabilities.  SNPs are called associated
if their MISA SNP BF is greater than $3.2$.  All SNPs that did not
pass the marginal screen step in MISA were declared not associated.
\end{longlist}
The first four are single SNP methods, while the last three are
multi-SNP methods that take into account the genetic parametrization
for each SNP.
\begin{figure}

\includegraphics{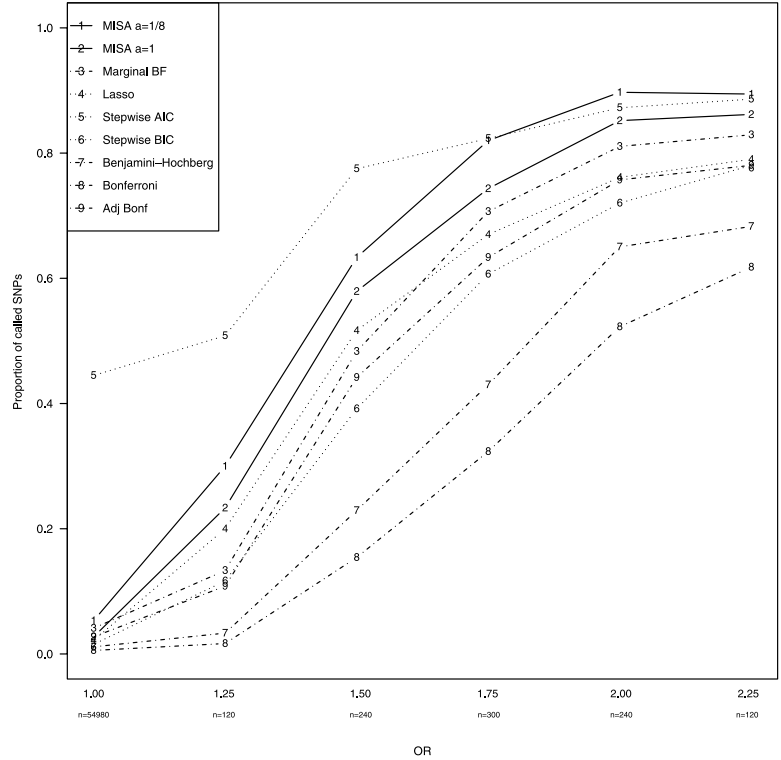}

\caption{True and false positive
rates of MISA versus alternative methods.}\label{power.ps}
\end{figure}

Figure~\ref{power.ps} shows the proportion of SNPs detected by each of
the methods as a function of the assumed true odds ratio.  Thus, at an
odds ratio of 1.00 we plot the proportion of SNPs that were falsely
declared associated by each of the methods.  While both Bonferroni and
Benjamini--Hochberg have the smallest false positive rates, they have
much lower power to detect true associations than any of the other
methods; the marginal BF has the highest power out of the three
marginal methods, and is comparable to lasso, a multi-SNP method.
Stepwise model selection using BIC has the lowest power of the multiple SNP
model selection procedures.  Stepwise logistic regression using AIC to
select a model, on the other hand, has high power to detect
associations, but an unacceptably high false positive rate ($44\%$).
With the exception of stepwise/AIC, the MISA methods have higher power
than the alternatives at all odds ratios (ORs) in the simulation, with
the gain in power most noticeable for the smaller ORs, those
encompassing the range 1.25--1.75 typically seen in practice
[\citep{FlinMack2009}].  This increase in power comes at the cost of
only a slight increase in the false positive rate.  Overall, MISA
using the default $\BB(1,S)$ prior distribution is able to detect 9\%
as many associations at the SNP level and 13\% as many at the gene
level than the marginal BF method used alone.  In addition, MISA is
able to detect 19\% as many true associations at the SNP level and
27\% as many at the gene level as the calibrated Bonferroni method
(the two methods have the same Type I error rate).

\subsection{Sensitivity to  hyperparameters}\label{section:simchoice}

We examined a range of parameters ($a$ and $b$) for the Beta-Binomial
prior distribution on model size (Table~\ref{fp.tp.table}) to assess
sensitivity of true positive and false positive rates.  In practice,
this may be done by reweighting the MCMC output using the new prior
distribution, without resorting to additional MCMC runs, as long as
high posterior probability models receive adequate support under both
prior distributions.

\begin{table}[b]
\tabcolsep=0pt
\caption{Estimated overall false and true positive rates with standard
errors and prior odds (PO) of association at the gene and SNP
levels. The values in bold characterize the method selected for
use in the analysis of the NCOCS ovarian cancer
example}\label{fp.tp.table}
\begin{tabular*}{\textwidth}{@{\extracolsep{4in minus 4in}}l@{ \ \ }c@{\hspace*{7pt}} ll llcl@{}}
\hline
\multicolumn{2}{@{}r}{\textbf{Method:}}&
\multicolumn{2}{c}{\textbf{True positive}}&
\multicolumn{2}{c}{\textbf{False positive}}&
\multicolumn{2}{c@{}}{\textbf{PO of assoc.}}
\\[-6pt]
&&
\multicolumn{2}{c}{\hrulefill}&
\multicolumn{2}{c}{\hrulefill}&
\multicolumn{2}{c@{}}{\hrulefill}
\\
& & \multicolumn{1}{@{}c}{\textbf{Gene (se)}} & \multicolumn{1}{c}{\textbf{SNP (se)}}
& \multicolumn{1}{c}{\textbf{Gene (se)}} & \multicolumn{1}{c}{\textbf{SNP (se)}}
& \multicolumn{1}{c}{\textbf{Global}} & \multicolumn{1}{c@{}}{\textbf{SNP}}
\\
\multicolumn{2}{@{}r}{$\bolds{n}$\textbf{:}}
& \multicolumn{1}{c}{\textbf{1020}}
& \multicolumn{1}{c}{\textbf{1020}}
& \multicolumn{1}{c}{\textbf{5546}}
& \multicolumn{1}{c}{\textbf{54980}} &  &
\\[3pt]
\multicolumn{2}{@{}l}{\textbf{MISA}}
\\[-6pt]
\multicolumn{2}{@{}c}{\hrulefill}
\\
 $\bolds{a}$&$\bolds{b}$  &&&&&&
\\
\hline
 1& $ \frac{1}{2}S$ & 0.77 (0.006) & 0.669 (0.007) & 0.128 (0.001) & 0.025 (0.0001) &2.00 & 0.04 \\
 $1/2$ &$\cdot$& 0.809 (0.005) & 0.704 (0.007) & 0.166 (0.001) & 0.031 (0.0001) & 0.74 & 0.020 \\
 $1/4$ &$\cdot$& 0.846 (0.004) & 0.729 (0.006) & 0.189 (0.001) & 0.041 (0.0002) & 0.32 & 0.009 \\
 $1/8$ &$\cdot$& 0.874 (0.003) & 0.739  (0.006) & 0.259 (0.001) & 0.048 (0.0002) & 0.15 & 0.005 \\
 $1/16$ &$\cdot$& 0.896 (0.003) & 0.746 (0.006) & 0.341 (0.001) & 0.065 (0.0003)& 0.07 & 0.002\\
 $1/32$ &$\cdot$& 0.904 (0.003) & 0.746 (0.006) & 0.437 (0.001) & 0.090 (0.0003) & 0.04 & 0.001\\
[5pt]
 1 & $S$& 0.784 (0.005) & 0.685 (0.007) & 0.150 (0.001) & 0.027 (0.0001) &1.00 & 0.020 \\
 $1/2$ &$\cdot$& 0.821 (0.005) & 0.716 (0.006) & 0.185 (0.001) & 0.035 (0.0001) & 0.42 & 0.009\\
 $1/4$ &$\cdot$& 0.855 (0.004) & 0.736 (0.006) & 0.207 (0.001) & 0.044 (0.0002) & 0.19 & 0.005\\
 \multicolumn{1}{@{}l}{$\mathbf{1/8}$}
& $\cdot$
& \multicolumn{1}{l}{$\mathbf{0.877\ (0.003)}$}
& \multicolumn{1}{l}{$\mathbf{0.743\ (0.006)}$}
& \multicolumn{1}{l}{$\mathbf{0.280\ (0.001)}$}
& \multicolumn{1}{l}{$\mathbf{0.053\ (0.0002)}$}
& \multicolumn{1}{l}{\hspace*{5pt}$\mathbf{0.09}$}
& \multicolumn{1}{l@{}}{$\mathbf{0.002}$}
\\
 $1/16$ &$\cdot$& 0.899 (0.003) & 0.746 (0.006) & 0.368 (0.001) & 0.073 (0.0003) & 0.04 & 0.001 \\
 $1/32$ &$\cdot$& 0.904 (0.003) & 0.746 (0.006) & 0.465 (0.001) & 0.098 (0.0004) & 0.02 & 0.001\\
[5pt]
 1 &$ \frac{3}{2}S$& 0.791 (0.005) & 0.696 (0.007) & 0.169 (0.001) & 0.029 (0.0001) & 0.67 & 0.01  \\
$1/2$ &$\cdot$& 0.825 (0.005) & 0.722 (0.006) & 0.190 (0.001) & 0.037 (0.0002)  & 0.29 & 0.006 \\
$1/4$ &$\cdot$& 0.855 (0.004) & 0.735 (0.006) & 0.222 (0.001) & 0.048 (0.0002) & 0.14 & 0.003 \\
$1/8$ &$\cdot$& 0.878 (0.003) & 0.744 (0.006) & 0.291 (0.001) & 0.057 (0.0002) & 0.07 & 0.002 \\
$1/16$ &$\cdot$&  0.898 (0.003) & 0.746 (0.006) & 0.377 (0.001) & 0.075 (0.0003) & 0.03 & 0.001 \\
$1/32$ &$\cdot$& 0.902 (0.003) & 0.746 (0.006) & 0.474 (0.001) & 0.099 (0.0004) & 0.02 & 0.0004 \\
[5pt]
\multicolumn{2}{@{}l}{Marg.  BF }  & 0.695 (0.007) & 0.627 (0.007) & 0.171 (0.001) & 0.041 (0.0002) &\multicolumn{1}{c}{--}&1.00\\
\multicolumn{2}{@{}l}{lasso } & 0.708 (0.007) &  0.607 (0.008)    & 0.158(0.001) &  0.022 (0.0001) &\multicolumn{1}{c}{--}&\multicolumn{1}{c@{}}{--}\\
\multicolumn{2}{@{}l}{Step. AIC}  & 0.993 (0.000) & 0.794 (0.005)& 0.969 (0.0001) & 0.445 (0.001) &\multicolumn{1}{c}{--}&\multicolumn{1}{c@{}}{--}\\
\multicolumn{2}{@{}l}{Step. BIC}  & 0.680 (0.007) & 0.547 (0.008)  & 0.122(0.001)& 0.015 (0.0001) &\multicolumn{1}{c}{--}&\multicolumn{1}{c@{}}{--}\\
\multicolumn{2}{@{}l}{BH} & 0.439 (0.008) & 0.419 (0.008)  & 0.013 (0.0001)& 0.011 (0.0001) &\multicolumn{1}{c}{--}&\multicolumn{1}{c@{}}{--}\\
\multicolumn{2}{@{}l}{Bonf.} & 0.337 (0.007) & 0.330 (0.008)  & 0.003 (0.00001) & 0.006 (0.00002) &\multicolumn{1}{c}{--}&\multicolumn{1}{c@{}}{--}\\
\multicolumn{2}{@{}l}{Adj. Bonf. 1} &  0.618 (0.007) &  0.574 (0.008) &   0.069 (0.0003) &  0.027 (0.0001) &\multicolumn{1}{c}{--}&\multicolumn{1}{c@{}}{--}\\
\multicolumn{2}{@{}l}{Adj. Bonf. 2} & 0.708 (0.007) & 0.644 (0.007)  & 0.184 (0.001) & 0.053 (0.0002) &\multicolumn{1}{c}{--}&\multicolumn{1}{c@{}}{--}\\
\hline
\end{tabular*}
\end{table}

Over the range of values for $(a,b)$, MISA has a higher gene and SNP
true positive rate than any of the other simpler procedures, with the
exception of Stepwise AIC. In general, decreasing $a$ leads to higher
true positive rates, but at the expense of higher false positive
rates.  The SNP false positive rate is modest, ranging from $0.025$ to
$0.099$, providing effective control of the experiment wide error
rate.  While these rates are higher than the false positive rates
under Bonferroni or Benjamini--Hochberg, eliminating a SNP from
consideration that truly is associated has a higher scientific cost
than continuing to collect data to confirm that a SNP is really a null
finding.  Because the NCOCS will follow up apparent associations,
a higher true positive rate with a modest increase in false positives
was preferable.

The hyperparameters $a=1/8$ and $b=S$, highlighted in bold in Table~\ref{fp.tp.table}, were selected for comparison with the default choice
($a=1$, $b=S$) in the analysis of the NCOCS data presented in the next
section. MISA using the $\BB(1/8,S)$ is able to detect 19\% as many
true associations at the SNP level and 26\% as many at the gene level
as the marginal BF method used alone.  In addition, MISA with the
$\BB(1/8,S)$ prior is
able to detect 14\% as many true associations at the SNP level and
24\% as many at the gene level as a calibrated Bonferroni method
(the two methods have the same Type I error rate).

\section{Ovarian cancer association analysis}
\label{sec:NCOCS}
In this section we describe a MISA candidate pathway analysis of data
from the ongoing NCOCS ovarian cancer case-control association study.
The NCOCS is a population based study that covers a 48 county region
of North Carolina [\citep{Schildkraut2008}].  Cases are between 20 and
74 years of age and were diagnosed with primary invasive or borderline
epithelial ovarian cancer after January 1, 1999.  Controls are
frequency matched to the cases by age and race and have no previous
diagnosis of ovarian cancer.  In the analysis we present, we focus on
self-reported Caucasians and a specific histological subtype of the
cancer, leaving us a total of 397 cases and 787 controls.  Because the
ovarian cancer results have not yet been published, we have anonomyzed
the pathway, the genes chosen to represent it and the IDs of the SNPs
tagging variation in those genes.  The pathway is comprised of 53
genes tagged by 508 tag SNPs.

All models fit in the screen and by MISA included the patient's age as
a design variable.  We used the modal approximation to fill in missing
SNP data.  We screened 508 SNPs using marginal Bayes factors,
retaining $S = 70$ SNPs that exceeded the threshold of 1 in favor of
an association.  Using the default hyperparameters $a = 1$ and $b=S$,
we ran two independent runs of the algorithm from independent starting
points for a total of $1.2$ million iterations---the point at which
the SNP marginal inclusion probabilities from the two independent runs
were determined to be in sufficiently close agreement.

\begin{figure}

\includegraphics{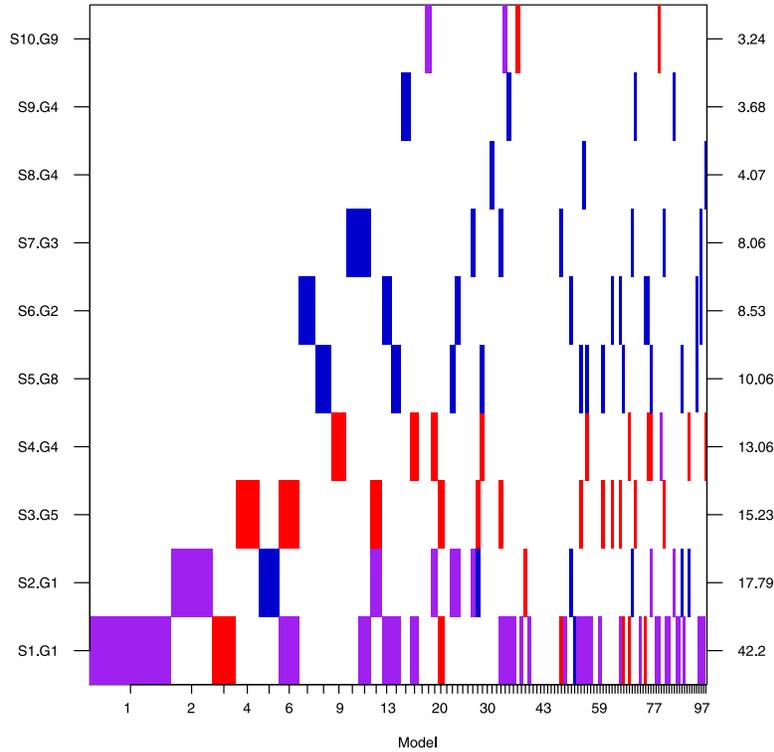}

\caption{Image plot of the SNP inclusion indicators for the SNPs with
marginal Bayes factors greater than 3.2 and the top 100 Models.  The
color of the inclusion block corresponds to the genetic
parametrization of the SNP in that model.  Purple corresponds to a
log-additive parametrization, red to a dominant parametrization and
blue to a recessive parametrization.  SNPs are ordered on the basis of
their marginal SNP Bayes factors which are plotted on the right axis
across from the SNP of interest.  Width of the column associated
with a model is proportional to its estimated model
probability.}\label{SNPinc}
\end{figure}

\begin{figure}

\includegraphics{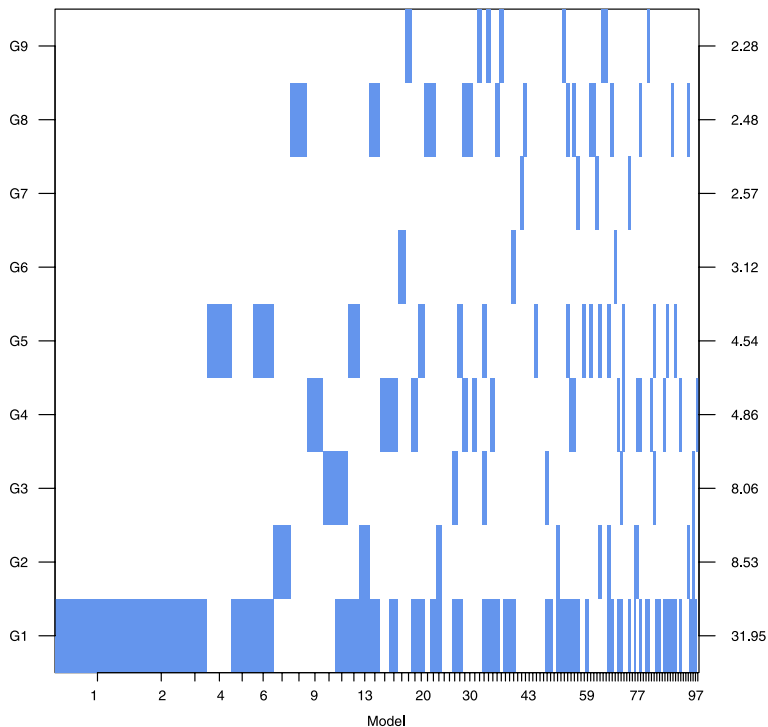}

\caption{Image plot of the gene inclusion indicators for the top 100
Models.  Genes are ordered based on their marginal gene Bayes
factors which are plotted on the right axis.  Columns correspond to
models and have width proportional to the estimated model
probability, models are plotted in descending order of posterior
support.  The color is chosen to be neutral since the genetic
parametrizations are not defined at the gene level.}
\label{GENEinc}
\end{figure}

On the basis of this analysis, we estimate a lower bound on the
pathway-wide Bayes factor  for association to be $\BF(H_A\dvtx H_0) =
7.67$  (which is also the posterior odds for this
prior).  This constitutes ``positive'' evidence in favor of an association
between the pathway and ovarian cancer based on Jeffreys' grades of
evidence and corresponds to a posterior probability that the pathway
is associated of roughly 0.89.
Figure~\ref{SNPinc} summarizes the associations of the ten
SNPs that had a SNP BF greater than 3.2, while Figure~\ref{GENEinc}
illustrates the nine genes that contained these SNPs and two others
that received comparable support.  SNPs and genes in the pathway are
denoted by a two-level name (e.g., S1 and G1) where the number
represents the rank of the SNP or gene by its respective Bayes factor.
These plots provide a graphical illustration of the top 100 models
$\Mg \in \M$ selected on the basis of their posterior model probabilities.
Models are ordered on the $x$-axis in descending probability and the
width of the column associated with a model is proportional to that
probability.  SNPs (Figure~\ref{SNPinc}) or genes (Figure~\ref{GENEinc}) are represented on the $y$-axis.  The presence of a SNP
or gene in a model is indicated by a colored block at the intersection
of the model's column and the SNP's or gene's row.  In Figure~\ref{SNPinc} the color of the block indicates the parametrization of
the SNP: purple for log-additive, blue for recessive and red for
dominant.  The ``checkerboard'' pattern  (as opposed to the presences
of more vertical bars) suggests substantial model uncertainty.

The top five models depicted in Figure~\ref{SNPinc} include only a
single SNP in addition to age at diagnosis (the design variable is
omitted in the figure as it is included in all models).  The top model
includes SNP S1 in gene G1 under the log-additive genetic
parametrization, which is estimated to have an odds ratio (OR) of
approximately 1.42 (the posterior mode).  The second ranked model
includes only SNP S2 in gene G1 under the log-additive genetic
parametrization with an estimated OR of 1.37.  Note that the study has
relatively low power to detect effects of this magnitude (Figure~\ref{power.ps}).

Figure~\ref{SNPinc} also illustrates that many of the top models
beyond the first five include multiple SNPs.  This suggests that if we
were to restrict our attention to single SNP models, we would
potentially lose substantial information regarding their joint
effects.  For example, model six is comprised of both SNP S3 from gene
G5 and SNP S1 from gene G1, while model 12 is comprised of both SNP S3
from gene G5 and SNP S2 from gene G2.  In both cases, SNP S3 is
included in models with a SNP from gene G1.  This may indicate that
not only are SNPs S1, S2 and S3 important as single effects in the
top four models, but that their combined effects may be of interest.
Note that, in cases where the disease variant is unmeasured but
``tagged,'' several tagged SNPs may be required to explain variation at
that locus.

The SNP Bayes factors of S1 (BF${}={}$42.2) and S2 (BF${}={}$17.8) provide
``strong evidence'' of changes in prior beliefs, however, the marginal
posterior probabilities of association with ovarian cancer are 0.38
and 0.20, respectively.  Figure~\ref{SNPinc} illustrates that when one
of SNP S1 or S2 is included in a model, the other is often not (at
least in the top 50 models). This trade off often arises when SNPs are
correlated (i.e., in high linkage disequilibrium).  In this case, $R^2$
is 0.5 suggesting fairly strong LD between SNPs S1 and S2, in which case
the joint inclusion probabilities are more meaningful than marginal
probabilities.  Both SNP 1 and SNP 2 are in gene G1 which has a gene
Bayes factor of 31.95 (Figure~\ref{GENEinc}) and posterior probability
of association of 0.58.  These probabilities need to be interpreted in
the context of model uncertainty; conditional on the pathway being
associated with ovarian cancer, the probability that gene G1 is
driving the association is 0.58$/$0.89${}={}$0.65.  However, there remains
substantial uncertainty regarding which genes and SNPs may explain it,
as the posterior mass is spread over competing models/hypotheses.  The
positive support for an association suggests the continuation of data
accrual to refine these posterior probabilities.

Gene G1 and other genes in Figure~\ref{GENEinc} highlight a caution
regarding the interpretation of Bayes factors as a measure of absolute
support with composite hypotheses.  The gene Bayes factor for G1 is
31.95, which is smaller than the SNP Bayes factors for S1 (42.2).  The
posterior probability that gene G1 is associated is based on summing
the probabilities of all models that include at least one SNP from
that gene (S1, S2 and S51), hence, the \textit{posterior probability} for
gene inclusion is always greater than or equal to the probability that
any one SNP is included (i.e., posterior probabilities observe a
monotonicity property with composite hypotheses).  Bayes factors (and
$p$-values) for composite hypotheses do not share this monotonicity
property [\citep{LaviSche1997}].  Bayes factors for comparing
composite hypotheses may be expressed as the ratio of the weighted
average (with respect to the prior distribution) of marginal
likelihoods conditional on the hypotheses, which may decrease the
evidence in favor of a composite hypothesis when a subset of the
individual hypotheses have low likelihood.  As mentioned in Section~\ref{sec2.1.4}, while Bayes factors do not provide a coherent measure of
\textit{absolute} support because of their nonmonotonicity property,
\citet{LaviSche1997} show that the log Bayes factor does provide a
coherent measure of how much the data \textit{change} the support for the
hypothesis (relative to the prior).  Hence, they do provide useful
summaries of changes in prior beliefs of association in large
association studies with many competing models/hypotheses.

\subsection{Sensitivity analysis}
In this section we consider sensitivity of the results in the NCOCS
study to the prior distribution on the models and to the method of
imputation.  The simulation study suggests that priors with smaller
values of $a$ may identify more associated SNPs.  We estimated that
the $\BB(1/8,S)$ prior distribution on model size has a false positive
rate comparable to the marginal BF method, but a much higher true
positive rate, in the scenarios we considered.  Full data imputation,
achieved by averaging over the distribution of missing SNPs, is
probabilistically correct, but computationally expensive.  Thus, if
the use of modal imputation provides an accurate approximation to BF
calculated using full imputation, the computational efficiency of MISA
can be greatly improved at small cost.

For purposes of this analysis, we used the set of unique models
identified by the EMC search with modal imputations and $a = 1$ and
calculated 3 additional sets of BFs.  First, we obtained marginal
likelihoods for each of these models using 100 imputed data sets with
missing SNPs filled in based on their estimated distribution.  Second,
we calculated BFs using the $\BB(1/8, S)$ and $\BB(1, S)$ prior
distributions using the marginal likelihoods under the full and modal
imputations.  We applied ANOVA to these four sets of BFs to compare
the effects of prior hyperparameters and imputation methods after
adjusting for SNP using the ranked SNP BFs.\footnote{Ranks that were used
as residuals on the log scale still exhibited strong departures from
normality.}

\begin{table}[b]
\caption{Analysis of variance for the ranked SNP Bayes factors
contrasting the prior hyperparameters (default $a=1$ versus $a=1/8$)
and method of imputation (full imputation with 100 data sets versus
a modal estimate of the missing genotypes) for the 70 SNPs in
the NCOCS pathway that passed the marginal screen} \label{tab:anova}
\begin{tabular*}{\textwidth}{@{\extracolsep{4in minus 4in}}ld{3.0} d{7.2}d{6.2}d{4.2}d{1.4}@{}}
\hline
& \multicolumn{1}{c}{\textbf{d.f.}}
& \multicolumn{1}{c}{\textbf{Sum Sq.}}
& \multicolumn{1}{c}{\textbf{Mean Sq.}}
& \multicolumn{1}{c}{$\bolds{F}$\textbf{-value}}
& \multicolumn{1}{c@{}}{\textbf{Pr(}$\bolds{{>}F}$\textbf{)}} \\
\hline
SNP             & 69    & 1635891.00    & 23708.57  & 208.04    & \multicolumn{1}{c@{}}{${<}2\times 10^{-16}$}\\
Prior           & 1     & 169641.66     & 169641.66 & 1488.60   & 0.0000 \\
Impute          & 1     & 134.41        & 134.41    & 1.18      & 0.28 \\
Prior:impute    & 1     & 53.16         & 53.16     & 0.47      & 0.50 \\
Residuals       & 207   & 23589.77      & 113.96 &  &  \\
\hline
\end{tabular*}
\end{table}

Table~\ref{tab:anova} shows that the method of imputation has no
significant effect on the ranking of SNP BFs.  This suggests that, for
purposes of model search and calculation of BFs, we may use the modal
imputed genotypes in place of full imputation, with significant
computational savings.  For purposes of parameter estimation, we
suggest that the use of full imputation using a subset of the top models
and top SNPs as using a plug-in approach for imputation is known to
underestimate uncertainty.

We anticipated that the prior distribution would have a significant
effect based on the higher true positive and false positive rates
estimated from the simulation study and by considering differences in
the prior odds. While Table~\ref{tab:anova} suggests that overall the
rankings are different between the two prior distributions, the top 20
SNPs have the same rank under each of the four methods, leading to no
qualitative differences in our conclusions about the top SNPs.  The
prior odds for any given SNP's inclusion in a model are $8$ times
lower under the $\BB(1/8,S)$ prior distribution than under to the
$\BB(1,S)$ prior distribution; the resulting SNP BFs are $2.8$ times
higher under the $\BB(1/8,S)$ prior distribution than those under the
$\BB(1,S)$ prior distribution.  As a result, eight more SNPs are above
the 3.2 threshold used by the NCOCS to determine SNPs worthy of
additional study.

\subsection{External validation and comparison}
To provide a basis of comparison, we applied the methods described in
the simulation study (Section~\ref{section:compare}) to the NCOCS
data. We omitted stepwise logistic regression using AIC because of its
poor operating characteristics.  The marginal FDR methods of
Bonferroni and Benjamini--Hochberg failed to identify any significant
SNPs.  Lasso, which accounts for correlation among SNPS, also failed
to identify any SNPS.  Stepwise logistic regression using BIC selected
a model with three of the top four SNPs identified by MISA---S1.G1,
S3.G5 and S4.G4---but failed to identify S2.G1, which has
correlation 0.71 with SNP S1.G1.  This highlights a problem with
selection methods that ignore model uncertainty.

The NCOCS proposed two SNPs---S10 and S14 in G9---for external
validation by the Ovarian Cancer Association Consortium (OCAC), a
large international multi-center consortium of ovarian cancer
case-control studies.  The decision to focus on these variants was
made on the basis of results from an earlier version of the NCOCS data set
and on the basis of the strong prior interest NCOCS researchers had in the
gene (and not on the basis of the analysis described above).  Under the
default $\BB(1,S)$ prior distribution, only SNP S10 in G9 exceeds the
3.2 threshold and the G9 BF is only 2.28. In contrast, under the
$\BB(1/8,S)$ prior distribution, both SNPs S10 and S14 (LD 0.62) in G9
have SNP BFs greater than 3.2 (8.70 and 5.99, respectively) and the
gene BF is 6.18. An additional three SNPs in the same gene were
proposed by another member of the consortium on the basis of
uncorrected $p$-values. Of the five SNPs proposed for validation, only
SNPs S10 and S14 were confirmed to be associated with serous invasive
ovarian cancer by OCAC [\citep{Schietal2009}].

\section{Discussion}\label{sec6}
In this paper we describe MISA, a natural framework for multi-level
inference with an implicit multiple comparisons correction for
hypothesis based association studies.  MISA allows one to quantify
evidence of association at three levels: global (e.g., pathway-wide),
gene and SNP, while also allowing for uncertainty in the genetic
parametrization of the markers.  We have evaluated MISA against
established, simple to implement and more commonly used methods and
demonstrated that our methodology does have higher power than these methods in
detecting associations in modestly powered candidate pathway
case-control studies.  The improvement in power is most noticeable
for odds ratios of modest (real world) magnitude and comes at the cost
of only a minimal increase in the false positive rate.  Like stepwise
logistic regression, lasso and logic regression, MISA improves upon
marginal, SNP-at-a-time methods by considering multivariate
adjusted associations.  By using model averaging, MISA improves upon
these multivariate methods that select a single model, which may miss
important SNPs because of LD structure.  These improvements have
concrete implications for data analysis: MISA identified SNPs in the
NCOCS data that were subsequently externally validated; none of the
less complex methods considered here highlighted these SNPs to be of
interest.  Currently, other top ranked SNPs in genes identified by
MISA are undergoing external validation.  Finally, we note that while
MISA was developed for binary outcomes in case-control studies, MISA
is readily adaptable to accommodate other forms of outcome variables
(e.g., quantitative traits or survival) that are naturally modeled
within a GLM framework.

\subsection*{Web resources}
The URL for the software for the methodology and simulations presented in this paper is as follows:
\url{http://www.isds.duke.edu/gbye/packages.html}.

\begin{supplement}
\stitle{Bayesian model search and multilevel inference
for SNP association studies: Supplementary materials}
\slink[doi]{10.1214/09-AOAS322SUPP}
\slink[url]{http://lib.stat.cmu.edu/aoas/322/supplement.pdf}
\sdatatype{.pdf}
\sdescription{In this supplement we provide details for:
(1) Derivation of the implied prior distribution on the regression
coefficients when AIC is used to approximate the marginal likelihood
in logistic regression, (2) Description of the marginal Bayes
factor screen used to reduce the number of SNPs in the MISA analysis,
(3) Details of how the simulated genetic data sets used in the power
analysis of MISA were created and information on the statistical software we developed for this purpose, and
(4) Location of the freely available software resources referred to in this and the parent document.}
\end{supplement}

\printaddresses

\end{document}